\documentclass[final,3p,times,twocolumn]{elsarticle}

\journal{Systems \& Control Letters}

\usepackage{graphicx}          
\usepackage{graphics}
\usepackage{subfigure}

\usepackage{amsmath,amssymb,amsfonts}
\DeclareMathAlphabet{\mathcal}{OMS}{cmsy}{m}{n}

\allowdisplaybreaks
\usepackage{hyperref}
\usepackage{cleveref}

\usepackage{color}
\usepackage{float}

\newtheorem{thm}{Theorem}
\newtheorem{lem}{Lemma}
\newtheorem{cor}{Corollary}
\newdefinition{rem}{Remark}
\newproof{pf}{Proof}

\newcommand{\icl}{\textcolor{black}}
\newcommand{\il}{\textcolor{black}}
\usepackage[prependcaption,colorinlistoftodos]{todonotes}

\begin{document}

\begin{frontmatter}


\author{Mengmou Li\fnref{cam}}
\ead{ml995@cam.ac.uk}
\author{Ioannis Lestas\fnref{cam}}\ead{icl20@cam.ac.uk}
\author{Li Qiu\fnref{hkust}}\ead{eeqiu@ust.hk}
 \address[cam]{Department of Engineering, University of Cambridge, Cambridge, CB2 1PZ, UK}
 \address[hkust]{Department of Electronic and Computer Engineering, The Hong Kong University of Science and Technology, Clear Water Bay, Kowloon, Hong Kong, China}

\title{Parallel Feedforward Compensation for Output Synchronization: Fully Distributed Control and Indefinite Laplacian}


%


\begin{abstract}
This work is associated with the use of parallel feedforward compensators (PFCs) for the problem of output synchronization over heterogeneous agents and the benefits this approach can provide. Specifically, it addresses the addition of stable PFCs on agents that interact with each other using diffusive couplings.
The value in the application of such PFC is twofold.
Firstly, it has been an issue that output synchronization among passivity-short systems requires global information for the design of controllers in the cases when initial conditions need to be taken into account, such as average consensus and distributed optimization. We show that a stable PFC can be designed to passivate a passivity-short system while its output asymptotically vanishes as its input tends to zero. As a result, output synchronization is achieved among these systems by fully distributed controls without altering the original consensus results.
Secondly, in the literature of output synchronization over signed weighted graphs, it is generally required that the graph Laplacian be positive semidefinite, i.e., $L \geq 0$ for undirected graphs or $L + L^T \geq 0$ for balanced directed graphs.
We show that the PFC serves as output feedback to the communication graph to enhance the robustness against negative weight edges. As a result, output synchronization is achieved over a signed weighted and balanced graph, even if the corresponding Laplacian is not positive semidefinite.
\end{abstract}




\end{frontmatter}


\section{Introduction}
Due to the need for information consensus in large-scale complex networks, and also its close connection with output regulation in control, output synchronization of heterogeneous systems has received intense research attention over the past two decades \cite{chopra2006passivity,wieland2011internal,wieland2013synchronous,qu2014modularized,isidori2014robust,zhu2016general,baldi2018output}. \cite{wieland2011internal} constructs a necessary and sufficient internal model principle for heterogeneous linear systems. \cite{wieland2013synchronous} derives necessary conditions for non-identical nonlinear systems. \cite{chopra2006passivity,qu2014modularized} approach this problem with passivity and passivity indices.  In addition, robustness \icl{issues 
are} also considered in \cite{isidori2014robust,zhu2016general,baldi2018output}.
It is worth mentioning that while most of the above works apply the static diffusive couplings of outputs along with some feedback controls, \cite{wieland2011internal} considers augmenting a system to form the dynamic couplings for synchronization. This idea is closely related to the parallel feedforward compensator.

The use of parallel feedforward compensator (PFC) technique dates back to the work \cite{bar1987parallel}, where a compensator is added in parallel to render a linear system almost strictly positive real (ASPR), motivated by the need to implement an adaptive controller on positive real plants.
Recently, \cite{kim2016design} provides a necessary and sufficient condition for the existence of a stable PFC to render a single-input-single-output (SISO) LTI system minimum phase.
There are also some recent applications of PFC in the literature. It is adopted in \cite{yamashita2020passivity} to generalize the primal-dual algorithm to a class of new algorithms that are convergent under general convex objective functions. The widely used proportional-derivative (PD) control in distributed networks \cite{lombana2014distributed, ma2019exact} can be viewed as an application of PFC on top of the proportional control. Consensus controllers for linear multi-agent systems in \cite{li2009consensus} can be interpreted by interaction laws augmented by PFC \cite{kim2016design}. Moreover, the modified proportional-integral (PI) algorithm addressed in \cite{kia2015distributed,li2020input} and the Zero-Gradient-Sum (ZGS) algorithm proposed in \cite{lu2012zero} for distributed optimization are unified via PFC (see \nameref{example 3}).
However, compared to the rich literature of feedback design (e.g., \cite{khalil1996nonlinear,sepulchre2012constructive}), there are fewer works on the design of PFC.
The feedforward design is also applied to state synchronization in \cite{kim2016design} and output synchronization in \cite{wieland2011internal} with fully distributed controllers but these works do not address the case when initial conditions need to be taken into account, e.g., average consensus and distributed optimization \cite{olfati2007consensus,lu2012zero,kia2015distributed}.
On the other hand, output synchronization of passivity-short systems requires global graph information to design coupling gains \cite{qu2014modularized,li2019consensus}.
Thus, designing a proper PFC to render fully distributed controllers without altering the final consensus results is worth addressing.

Synchronization over signed weighted graphs has also been widely studied in recent years, see, e.g., \cite{altafini2012consensus,zelazo2015robustness,chen2016definiteness}. A negative weight edge in a Laplacian matrix can represent the antagonistic interaction between two agents. Such a matrix is called a \textit{signed weighted Laplacian}.
Some works also focus on the study of inertias of signed weighted Laplacians, i.e., the number of positive, negative, and zero eigenvalues \cite{bronski2014spectral,chen2020spectral}.
\cite{zelazo2014definiteness} characterizes the positive semidefiniteness of signed weighted Laplacians in terms of effective resistance, while \cite{chen2016definiteness} gives a passivity-based interpretation. The research is extended to directed graphs in \cite{mukherjee2018robustness}. Negative weight edges are also seen in power systems \cite{song2017network,ding2017impact}, which corresponds to the negative reactance or large angle differences across certain transmission lines in a power network.
However, the aforementioned works require positive semidefiniteness on the graph Laplacian ($L \geq 0$, or $L + L^T \geq 0$) to guarantee convergence. Output synchronization with indefinite Laplacians has not thus far been dealt with in the literature.

The contributions of this paper are on the design and applications of parallel feedforward compensators in a feedback interconnection and in output synchronization in particular. More specifically: \\
\text{1}. We show that a passive PFC can be added to a passive feedback interconnection and preserve the passivity of the overall closed-loop system. Such a PFC with an excess of passivity can provide additional negative energy to guarantee asymptotic convergence for lossless passive systems.  Furthermore, in the problem of output synchronization, it leads to decentralized control policies, where the consensus objective is not affected.\\
\text{2}. We propose systematic conditions in the frequency domain for designing local PFCs to passivate a class of passivity-short systems called input feedforward passive systems. \\
\text{3}. In the problem of output synchronization with signed weighted graphs, we show that the positive semidefinite requirement for graph Laplacian is not necessary with an addition of PFC. Using proportional-derivative (PD) controllers derived from the simplest form of PFC suffices to guarantee convergence under indefinite Laplacians.

The rest of this paper is organized as follows. In \Cref{section: preliminaries}, some preliminaries on graph theory, passivity, and feedback interconnection are given. In \Cref{section: PFC}, the addition of PFC to the feedback interconnection of systems is addressed via passivity and the problems to be investigated are formulated. In \Cref{section: pfc on OS agent,section: pfc on OS graphs}, the effects of PFC on agent dynamics and the underlying graphs in output synchronization are addressed along with some illustrative examples. Finally, the work is concluded in \Cref{section: conclusion}.

\section{Preliminaries}\label{section: preliminaries}
Notation: $\mathbb{R}$ and $\mathbb{C}$ represent the sets of real and complex numbers, respectively.
$\text{Re}[\cdot]$ and $\text{Im}[\cdot]$ represent the real part and imaginary part of a complex number (matrix), respectively.
$\| \cdot \|$ denotes the matrix induced 2-norm.
Let $\bigoplus$ denote the direct sum of matrices, i.e., $\bigoplus_{i = 1}^{N} A_i := \text{diag} \{A_1,\ldots,A_N \}$. $I_N$ denotes the $N \times N$ identity matrix, whose subscript can be omitted if its size is clear from the context.
$\mathbf{1}_N$ denotes the all-one vector in $\mathbb{R}^{N}$. $\mathbf{0}$ denotes a zero matrix of appropriate size. The conjugate and the conjugate transpose of a complex square matrix $A$ is denoted by $A^*$ and $A^{H}$, respectively.

\subsection{Graph theory and signed weighted graphs}
The communication of agents is represented by a directed graph $\mathcal{G} = (\mathcal{N},\mathcal{E}, \mathcal{A})$, where $\mathcal{N} = \{1,\ldots,N\}$ is the node set, $\mathcal{E}\subset \mathcal{N}\times\mathcal{N}$ is the edge set and $\mathcal{A}$ is the adjacency matrix.
The edge $(k,i) \in \mathcal{E}$ means that agent $i$ can obtain information from agent $k$.
We assume that there is no self-loop in $\mathcal{G}$, i.e., $(i,i) \notin \mathcal{E}$.
The weights of edges are assigned by $\mathcal{A} = [a_{ik}]$, where $a_{ik} \neq 0$ if $(k,i) \in \mathcal{E}$, and $a_{ik} = 0$, otherwise. In this work, $a_{ik}$ is not restricted to be nonnegative. $\mathcal{G}$ is said to be a \textit{signed weighted graph} if for some $(i,k)$, $a_{ik} < 0$. The corresponding $(i,k)$ is called \textit{negative weight edge}.
%
$\mathcal{G}$ is said to be \textit{strongly connected} if there exists a sequence of successive distinct edges between any two agents.
The graph $\mathcal{G}$ is said to be \textit{weight-balanced} if $\forall i \in \mathcal{N}$, the in-degree $d_{in}^{i} = \sum_{k=1}^{N} a_{ik}$ equals to out-degree $d_{out}^{i} = \sum_{k=1}^{N} a_{ki}$.
The Laplacian matrix of $\mathcal{G}$ is defined as $L = \text{diag}\{d_{in}^{1},\ldots, d_{in}^{N} \} - \mathcal{A}$. When $\mathcal{G}$ is weight-balanced, it satisfies that $\mathbf{1}_{N}^T L = \mathbf{0}$ and $L \mathbf{1}_{N} = \mathbf{0}$.

\subsection{Passivity and positive real transfer functions}
Consider a nonlinear system $\Sigma$ as follows
\begin{equation}\label{eq: nonlinear system state-space}
\begin{array}{rl}
\Sigma:
	\dot{x} = f(x, u), \quad y = h(x, u)
\end{array}
\end{equation}
where $f: \mathbb{R}^{p} \times \mathbb{R}^{m} \rightarrow \mathbb{R}^{p}$ is locally Lipschitz, $h: \mathbb{R}^{p} \times \mathbb{R}^{m} \rightarrow \mathbb{R}^{m}$ is continuous.
System $\Sigma$ is said to be \textit{passive} if there exists a nonnegative and differentiable real-valued function of the states $V(x)$, called the storage function, such that
	$\dot{V} \leq u^T y.$
It is said to be \textit{strictly passive} if $\dot{V} \leq u^Ty - \psi(x)$ for some positive definite function $\psi (x)$.
System $\Sigma$ is said to be \textit{input feedforward passive} (IFP) if $\dot{V} \leq u^T y - \nu u^T u$ for some $\nu \in \mathbb{R}$ and we denote this system as IFP($\nu$). When $\nu \leq 0$, the system is said to be \textit{passivity-short}.
System $\Sigma$ is said to be \textit{output feedback passive} (OFP) if $\dot{V} \leq u^T y - \rho y^T y$ for some $\rho \in \mathbb{R}$, denoted by OFP($\rho$).
\icl{Whenever we refer to a passive, IFP or OFP system, we will also assume its storage function is positive definite and radially unbounded. These are additional properties needed for the stability analysis in the paper\footnote{\icl{It should be noted that Lemma \ref{lem:OFP} and Theorem \ref{theorem: pfc preserves passivity} are also valid without these assumptions on the storage function.}}.}
For a memoryless function $y = h(u)$, $h: \mathbb{R}^{m} \rightarrow \mathbb{R}^{m}$ the same definitions hold for passivity, IFP, OFP respectively, but with the storage function $V$ set to zero.

A $m \times m$ proper rational transfer function matrix $H(s)$ is called \textit{positive real} if \\
1.~The poles of all elements of $H(s)$ are in $\{ s | \text{Re} [s] \leq 0 \}$,\\
2.~for all real $\omega$, where $j \omega$ is not a pole of any element of $H(s)$, $H(j \omega) + H^T(-j \omega) \geq 0$,\\
3.~Any pure imaginary pole $j \omega_{0}$ of any element of $H(s)$ is a simple pole, and the residue matrix $\underset{s \rightarrow j \omega_0}{\lim} (s - j \omega_0) H(s)$ is positive semidefinite Hermitian.\\
$H(s)$ is called \textit{strictly positive real} if $H(s - \varepsilon)$ is positive real for some $\varepsilon > 0$.
A minimal state space representation of the transfer function matrix $H(s)$ is passive if and only if $H(s)$ is positive real. It is strictly passive if $H(s)$ is strictly positive real. $H(s)$ is IFP($\nu$) if the inequality in the second condition is replaced by $H(j \omega) + H^T(-j \omega) \geq 2 \nu I$ for a bounded constant $\nu \in \mathbb{R}$.
\subsection{Feedback Interconnection}
\begin{figure}[htbp]
	\centering
	\includegraphics[width = .6\linewidth]{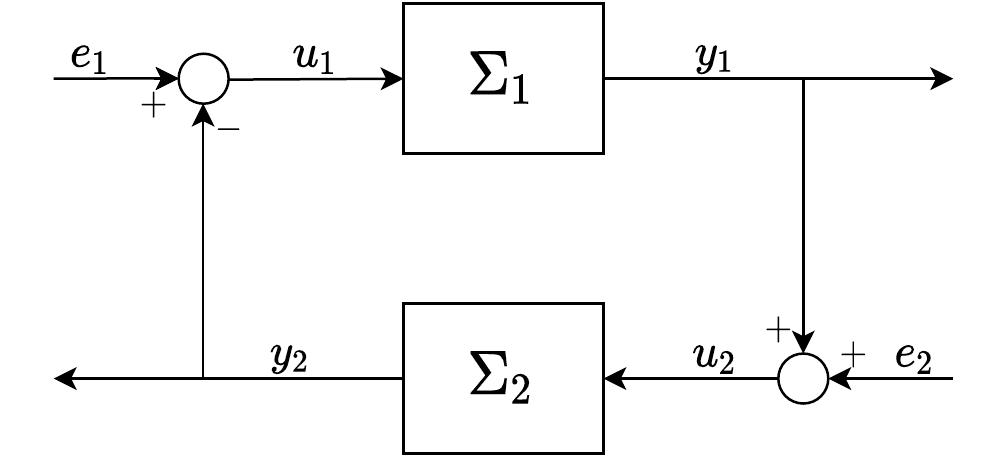}
	\caption{Feedback interconnection of two subsystems $\Sigma_1$ and $\Sigma_2$.}
	\label{fig: feedback interconnection}
\end{figure}

Consider two subsystems $\Sigma_1$ and $\Sigma_2$ represented by the nonlinear state space model
\begin{align}\label{eq: subsystems representation}
\Sigma_i :
\begin{array}{rl}
	\dot{x}_i =  f_i(x_i, u_i),~\quad
	y_i =  h_i(x_i, u_i)
\end{array}
\end{align}
with $x_i \in \mathbb{R}^{p_i}$, and $u_i,~ y_i \in \mathbb{R}^{m}$, $i = 1, 2$.
The negative feedback interconnection of the two subsystems as shown in \Cref{fig: feedback interconnection} results from the interaction rule
\begin{equation}\label{eq: feedback interconnection rule}
	u_1 = e_1 - y_2, \quad u_2 = e_2 + y_1.
\end{equation}
The closed-loop system is called \textit{well-posed} if there exists a unique solution which is causally dependent on inputs \cite{vidyasagar1981input}.  In this work, we assume that all the closed-loop systems considered are well-posed.
It is well-known that the feedback interconnection of two passive systems $\Sigma_1$, $\Sigma_2$ in \Cref{fig: feedback interconnection} is passive from $e: = \left[ \begin{smallmatrix} e_1\\ e_2 \end{smallmatrix} \right]$ to $y: = \left[\begin{smallmatrix} y_1\\ y_2 \end{smallmatrix} \right]$ if the closed-loop system is well-posed \cite{khalil1996nonlinear}.


\begin{lem}[OFP closed-loop system]\label{lem:OFP}
Suppose $\Sigma_1$ is \textup{OFP($\rho$)}, $\Sigma_2$ is \textsc{IFP($\nu$)}, and let $e_2 = \mathbf{0}$. Then the closed-loop system in \Cref{fig: feedback interconnection} is \textup{OFP($\rho + \nu$)} from $e_1$ to $y_1$.
\end{lem}
\begin{pf}
By definition, there exists a storage function $V_1 \geq 0$ for $\Sigma_1$ such that $\dot{V}_1 \leq y_1^T u_1 - \rho y_1^T y_1$, and a storage function $V_2 \geq 0$ for $\Sigma_2$ such that $\dot{V}_2 \leq y_2^T u_2 - \nu u_2^T u_2$. Let the overall storage function be $V = V_1 + V_2$. Substituting $e_2 = \mathbf{0}$ in \eqref{eq: feedback interconnection rule}, we have $\dot{V}_1 + \dot{V}_2 \leq y_1^T (e_1 - y_2) - \rho y_1^T y_1 + y_2^T y_1 - \nu y_1^T y_1 = y_1^T e_1 - (\rho + \nu) y_1^T y_1$.
\end{pf}

\section{Parallel Feedforward Compensator and Problem Formulation}\label{section: PFC}
In this section, we introduce the addition of a PFC to the closed-loop system of $\Sigma_1$, $\Sigma_2$ and then formulate the problems to be investigated.
\subsection{Parallel Feedforward Compensator}
\begin{figure}[H]
	\centering
	\includegraphics[width= .6\linewidth]{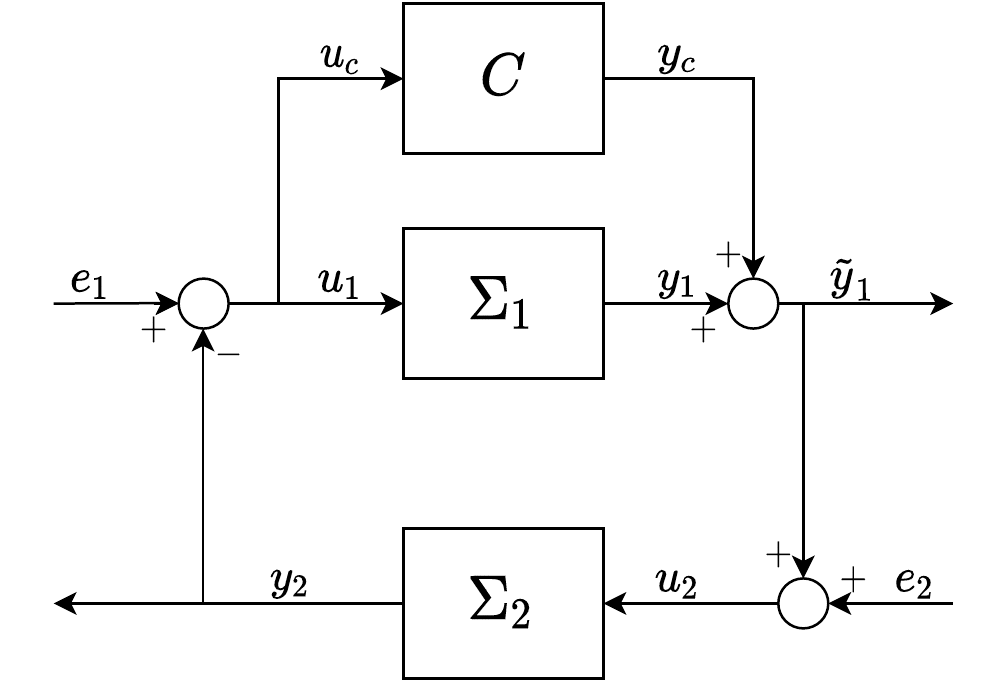}
	\caption{Parallel feedforward compensator $C$ added to the feedback interconnection of $\Sigma_1$ and $\Sigma_2$.}
	\label{fig: PFC to FI}
\end{figure}

A stable compensator $C$ is added in parallel to $\Sigma_1$, as shown in \Cref{fig: PFC to FI}.

The \textit{stable parallel feedforward compensator} $C$ is designed to be a system represented by
\begin{align}\label{eq: pfc state model}
		\dot{x}_{c} = f_{c} (x_c, u_c), \quad
		y_c = h_{c} (x_c, u_c)
\end{align}
where $f_c(\mathbf{0},\mathbf{0}) = \mathbf{0}$, $h_c(\mathbf{0}, \mathbf{0}) = \mathbf{0}$ and $x_c \rightarrow \mathbf{0}$, $y_c \rightarrow \mathbf{0}$ when $u_c \rightarrow \mathbf{0}$. Note that a common choice of such PFC can be a strictly passive system.
Differently from \cite{xia2018control}, where a static input-output transform matrix $M$ is introduced to passivate the system, we focus only on the feedforward operation but with extra internal dynamics $\dot{x}_{c} = f_{c} (x_c, u_c)$ that can affect the system performance.
The addition of such PFC is represented by the interconnection rule
\begin{equation}\label{eq: pfc interconnection rule}
\begin{array}{rlrl}
u_c = & u_1, & \tilde{y}_1 = y_1 + y_c,\\
u_1 = & e_1 -y_2, & u_2 = e_2 + \tilde{y}_1
\end{array}
\end{equation}
which preserves passivity of the closed-loop system if $C$ is passive.

\begin{thm}\label{theorem: pfc preserves passivity}
Suppose that the closed-loop system \eqref{eq: subsystems representation}, \eqref{eq: feedback interconnection rule} is passive from $e = \left[ \begin{smallmatrix} e_1\\ e_2 \end{smallmatrix} \right]$ to $y = \left[ \begin{smallmatrix} y_1\\ y_2 \end{smallmatrix} \right]$, and compensator $C$ is passive from $u_c$ to $y_c$. Then the closed-loop system  \eqref{eq: subsystems representation}, \eqref{eq: pfc state model} and \eqref{eq: pfc interconnection rule}, as shown in \Cref{fig: PFC to FI}, preserves passivity from $e = \left[ \begin{smallmatrix} e_1\\ e_2 \end{smallmatrix} \right]$ to $\tilde{y} := \left[ \begin{smallmatrix} \tilde{y}_1 \\ y_2 \end{smallmatrix} \right]$.
\end{thm}
\begin{pf}
By definition, there exist storage functions $V,~V_c \geq 0$ such that $\dot{V} \leq y_1^T e_1 + y_2 ^T e_2$, $\dot{V}_c \leq y_c^T u_c$ respectively. Here, with an abuse of notation, we notice that $e_2$ in \eqref{eq: feedback interconnection rule} is replaced by $e_2 + y_c$ in \eqref{eq: pfc interconnection rule}.
Then, let $S := V+ V_c \geq 0$. It follows from \eqref{eq: pfc interconnection rule} that $\dot{S} \leq y_1^T e_1 + y_2^T (y_c + e_2) + y_c^T (e_1 - y_2) = \tilde{y}_1^T e_1 + y_2^T e_2$.
\end{pf}
This theorem states that we can safely incorporate a compensator in parallel to a passive closed-loop system without affecting its stability. In addition, if $e_1 - y_2 \rightarrow \mathbf{0}$, then $\tilde{y}_1 \rightarrow y_1$. In other words, the PFC does not contribute to the final value of $y_1$.
Note that it only requires passivity of the closed-loop system but does not impose passivity requirement for $\Sigma_1$ or $\Sigma_2$ specifically. If $\Sigma_1$ is passive, then the passivity of the system follows directly from the fact that the parallel connection of the two systems $\Sigma_1$ and $C$ is passive.
\Cref{theorem: pfc preserves passivity} is useful when the system is only lossless passive but not asymptotically stable. A PFC can be added providing strictness of negative to guarantee asymptotic convergence results.


\subsection*{Example 1}\label{example 1}
To illustrate \Cref{theorem: pfc preserves passivity}, we consider the negative feedback interconnection rule \eqref{eq: feedback interconnection rule} of the two subsystems
\begin{align*}
\begin{array}{rl}
 \Sigma_i: \quad \dot{x}_i = u_i, \quad y_i = h_i(x_i), \quad x_i,~y_i \in \mathbb{R},~ i = 1, 2,
\end{array}
\end{align*}
where $h_i(x_i)$ is passive and radially unbounded, i.e., $h_i(x_i) \rightarrow \infty$ if $\|x_i\| \rightarrow \infty$. The overall system is
\begin{align}\label{eq:example 1 colsed-loop}
\dot{x}_1 = h_2(x_2); \quad \dot{x}_2 = - h_1(x_1)
\end{align}
lossless passive with the storage function $S(x) = \sum_{i = 1}^{2} \int_{0}^{x_i} h_i(\sigma) d\sigma \geq 0$. We add the PFC:
$
	\dot{x}_c = - x_c + u_c, \quad y_c = x_c
$
using the interconnection law \eqref{eq: pfc interconnection rule}, as depicted in \Cref{fig: PFC to FI}. The overall system becomes
\begin{align}\label{eq:example 1 colsed-loop with PFC}
\dot{x}_1 \hspace{-1mm} = \hspace{-1mm} h_2(x_2); \quad \dot{x}_2 \hspace{-1mm} = \hspace{-1mm} -  h_1(x_1)  \hspace{-0.5mm} -  \hspace{-0.5mm} x_c;  \quad \dot{x}_c \hspace{-1mm} = \hspace{-1mm} - x_c \hspace{-0.5mm}  + \hspace{-0.5mm} h_2(x_2).
\end{align}
Select the Lyapunov function candidate $V(x) = S(x) + \frac{1}{2} x_c^2$, which is radially unbounded and its time derivative is $\dot{V} = \tilde{y}_1 u_1 + y_2 u_2 - x_c^2 = - x_c^2 \leq 0$.  Invoking LaSalle's invariance principle, we have convergence to an invariant set $F$ where $\dot{V} = 0$. The latter implies that in set $F$ we have $x_c=0$ and therefore \icl{$h_2(x_2)=0$, which implies $x_2=0$ and hence $h_1(x_1)=0$}. Hence, set $F$ is the equilibrium point of the feedback interconnection of $\Sigma_1$ and $\Sigma_2$. \Cref{fig: example 1} below shows the stabilization of the interconnected system when $h_i(x_i) = x_i^3$.
\begin{figure}[htbp]
\centering
\subfigure[Trajectories of the lossless passive system \eqref{eq:example 1 colsed-loop}.]{\includegraphics[width = 0.48\linewidth]{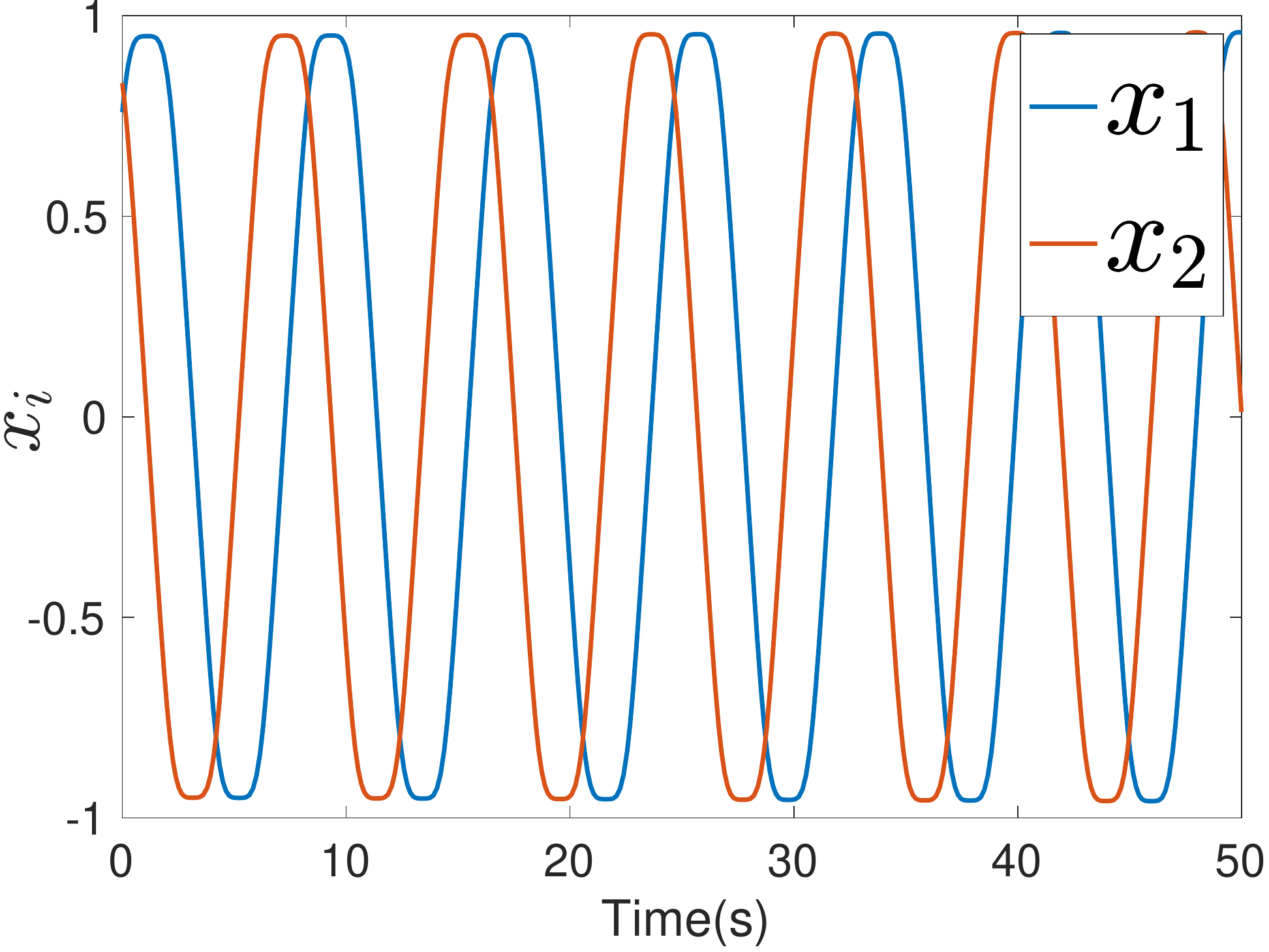}}
\subfigure[Trajectories of \eqref{eq:example 1 colsed-loop with PFC}, the lossless passive system with PFC.]{\includegraphics[width = 0.48\linewidth]{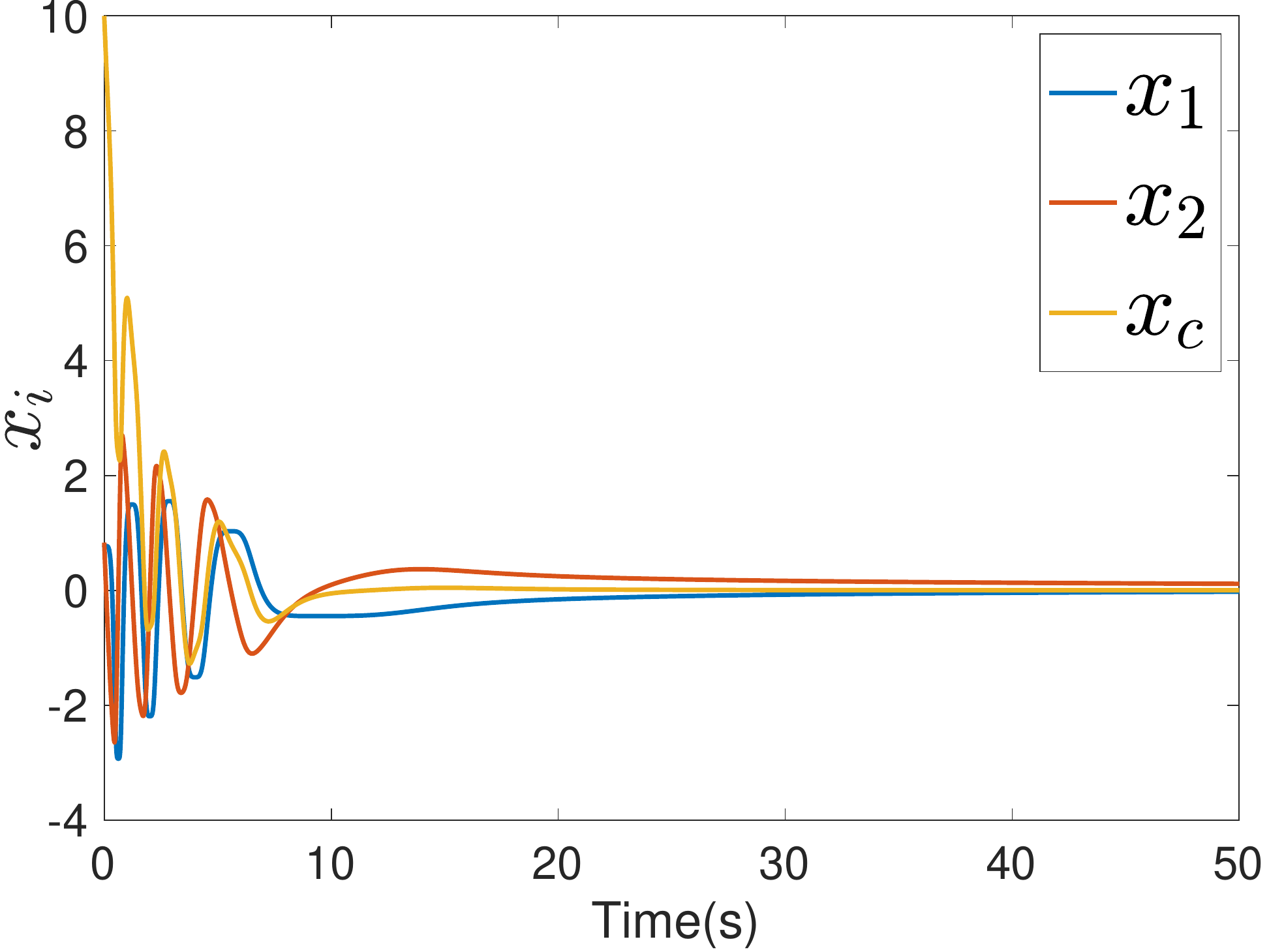}}
\caption{Stabilization of the lossless passive system \eqref{eq:example 1 colsed-loop} when $h_i(x_i) = x_i^3$.}
\label{fig: example 1}
\end{figure}

\subsection{Problem Formulation}
We will apply the framework in \Cref{fig: PFC to FI} to output synchronization hereafter.
Specifically, let $e_1 = \mathbf{0}$ and $e_2 = \mathbf{0}$ in \Cref{fig: PFC to FI}.
On the parallel augmented side, $\Sigma_1 = P := \bigoplus_{i =1}^{N} P_i$ represents the agent dynamics.
It consists of $N$ possibly heterogeneous agents and the input and output of each agent $P_i$ satisfy $u_{1i}, ~ y_{1i} \in \mathbb{R}^{m}$.
\il{Each local system $P_i$ has state vector $x_{1i}\in\mathbb{R}^{n_i}$, and is assumed to be affine nonlinear, i.e. it is given by:}
	\begin{align}\label{eq: affine nonlinear subsystem}
	\begin{array}{rl}
		\dot{x}_{1i} = f_{1i}(x_{1i}) + g_{1i}(x_{1i}) u_{1i}, \quad y_{1i} = h_{1i}(x_{1i}).
	\end{array}
	\end{align}

\begin{figure}[htbp]
	\centering
	\includegraphics[width = 0.6\linewidth]{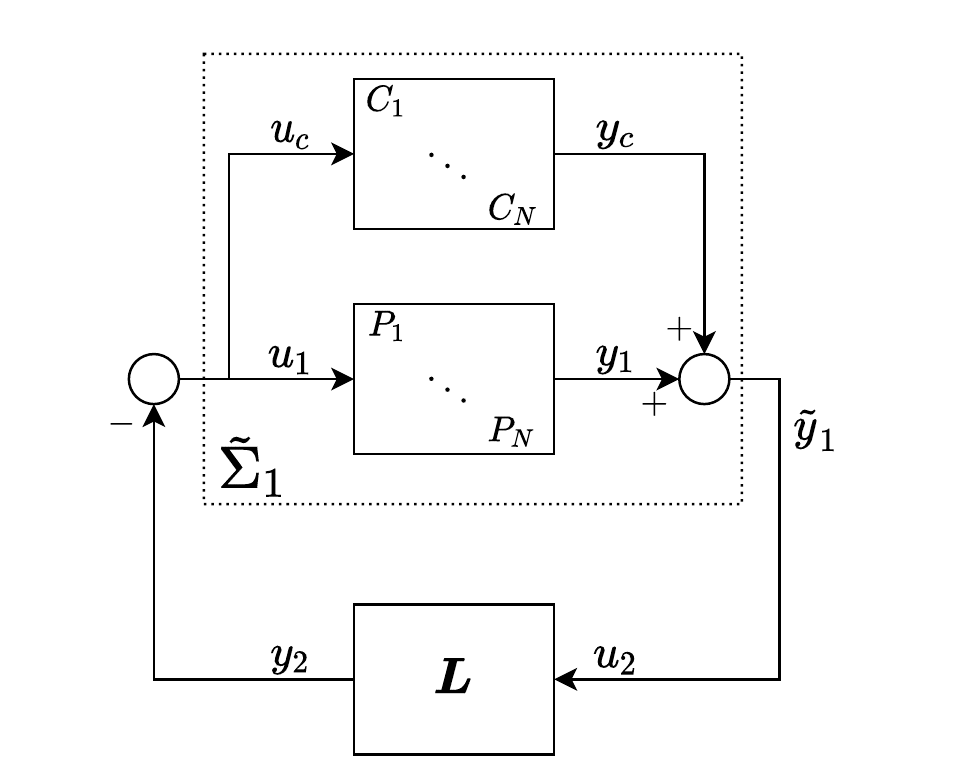}
	\caption{Output synchronization with PFC, where $\boldsymbol{L} = L \otimes I_m$, and $\tilde{\Sigma}_1$ represents the parallel interconnection system of $P$ and $C$.}
	\label{fig: PFCtoOS}
\end{figure}

On the other side, $\Sigma_2 = \boldsymbol{L} := L \otimes I_m$, which represents the interactions between agents. In other words, $\Sigma_2$ is the diffusive couplings among agents
\begin{equation}\label{eq: sigma_2 takes laplacian}
	y _2 = \boldsymbol{L} u_2.
\end{equation}
In the context of output synchronization, outputs of dynamics $P_i$ aim to reach consensus, i.e.,
\begin{align}\label{eq: output synchronization criteria}
\begin{array}{rl}
\underset{t \rightarrow \infty}{\lim} |y_{1i} -  y_{1k}| = 0, \quad \forall~i,~k \in \mathcal{N}.
\end{array}
\end{align}
The PFC is designed as $C := \bigoplus_{i = 1}^{N} C_i$, where $C_i$ is a local compensator added to agent $P_i$. The diagram is depicted by \Cref{fig: PFCtoOS}.
Let us adopt a stable compensator $C_i$ for each subsystems such that
\begin{equation}
y_{ci} \rightarrow \mathbf{0} \text{ when } u_{ci} \rightarrow \mathbf{0}.
\label{eq:PFC_stable}
\end{equation}
 We aim to discover the benefits of using the PFC while solving the following design problems.\\
\text{1}.~How to design a PFC to render fully distributed control when $P$ is passivity-short?\\
\text{2}.~How to design a PFC to guarantee convergence when the graph $\mathcal{G}$ is signed weighted?\\
We will address these problems in the following sections.
\section{Feedforward Compensation on Agent Dynamics}\label{section: pfc on OS agent}
Let us focus on the effects of PFC on $P$ in this section.
We adopt a linear system ($A_c,B_c,C_c,D_c$) as the compensator $C$.
Firstly, note that synchronization among passive systems is achieved through fully distributed diffusive couplings.
\begin{lem}[\hspace{1sp}\cite{chopra2006passivity}]\label{lem:fully distributed passive synchronization}
Let system $\Sigma_1 = P := \bigoplus_{i = 1}^{N} P_i$ be passive from $u_1$ to $y_1$, and $\Sigma_2 = \boldsymbol{L}$ in \Cref{fig: feedback interconnection}, and the graph is weight-balanced and strongly connected without negative weight edges. Then output synchronization is achieved.
\end{lem}
It is obvious that the storage function $V$ of $P$ satisfies $\dot{V} \leq y_1^T u_1 = - u_1^T \boldsymbol{L} u_1 \leq 0$. Applying LaSalle's invariance principle we can obtain the above result.
\Cref{lem:fully distributed passive synchronization} is said to be fully distributed because no global information is needed to design coupling controllers.
However, many stable systems are not passive, for example, a stable SISO linear plant whose Nyquist plot is not within the closed right-half complex plane is not passive. Then, we cannot apply \Cref{lem:fully distributed passive synchronization} to design fully distributed coupling controllers without shifting consensus results \cite{qu2014modularized,li2019consensus}.
Nevertheless, if the Nyquist trajectory does not go to infinity, i.e., it is input feedforward passive (IFP), then, it is always possible to drag it back to the right-half plane using a stable PFC. Therefore, we assume that each system $P_i$ is IFP, which applies to a wider class of Lyapunov stable systems and includes passivity as a special case.

Systems being IFP naturally implies that the consensus criterion \eqref{eq: output synchronization criteria} is achievable \cite{li2019consensus}.
The work \cite{bao2007process} adopts the same system dynamics and reshapes the output to passivate any Lyapunov stable system. However, by doing so, $y_{c}$ may not converge to zero and deviate from the goal to synchronize the original outputs.

\subsection{Passivation of linear input feedforward passive systems}
We give a straightforward method in the frequency domain for designing each compensator $C_i$ to passivate local linear IFP $P_i$.

Consider the MIMO linear system represented by the transfer function matrix $P_i(s)$. Then it can be decomposed into partial fractions as
\begin{align}\label{eq: transfer function mimo}
\begin{array}{rl}
\displaystyle P_i(s) = \sum^{n}_{l = 1} \left( \frac{R_{l1}}{s + d_l} + \cdots + \frac{R_{lk_{l}}}{(s + d_l)^{k_l}} \right)
\end{array}
\end{align}
where $R_{l1}, \ldots, R_{l k_{l}} \in \mathbb{R}^{m\times m}$.
Here, \eqref{eq: transfer function mimo} can represent a general class of transfer function matrices, which can have complex or multiple poles.
It is required that $\text{Re}[d_l] \geq 0$, $\forall l$, and any poles on the imaginary axis are simple to guarantee Lyapunov stability. In addition, for $P_i(s)$ to be IFP, the corresponding $R_l$ for any pure imaginary pole $d_l = j\omega_l$ is real and positive semidefinite.

It is worth mentioning that $\frac{A}{s + d}$ is passive for any $A > 0$ if $\text{Re} [d] \geq 0$. Then, we can use this term to passivate passivity-short systems.
\begin{thm}\label{theorem: mimo passivation}
	The \textsc{MIMO} \textsc{IFP} system given by \eqref{eq: transfer function mimo} with $R_{l1} = R_{l1}^T$ for all $l$, is input feedforward passivated by $C_i(s) = \sum_{l = 1}^{n} \frac{A_l}{s + d_l}$, where $A_l$ is a real symmetric matrix which satisfies that if $\text{Re}[d_l] = 0$, $A_l = \mathbf{0}$; and if $\text{Re}[d_l] \neq 0$,
\begin{equation}\label{eq: PFC on MIMO}
 A_l \geq  \left(  \frac{\left\| \text{Im}[d_l] \text{Im}[R_{l1}] \right\|}{\text{Re}[d_l]}  \hspace{-1mm}  +   \hspace{-1mm} \sum_{k = 1}^{k_l} \frac{  \left\| {R_{lk} + R_{lk}^{H}} \right\| + \left\| {R_{lk}^{H} - R_{lk}} \right\| }{{ 2 \text{Re}[d_l]^{k-1}}} \right) I.
	\end{equation}
\end{thm}
\begin{pf}
When $k_l = 1$, we let $G(s) = \frac{R_{l1}}{s + d_l} + \frac{R_{l1}^*}{s + d_l^*} + \frac{A_{l 1} }{s + d_l} + \frac{A_{l 1} }{s + d_l^*}$,. Note that $d_l = d_l^*$ and $R_{l1} = R_{l1}^*$ if $d_l \in \mathbb{R}$.  By $R_{l1} = R_{l1}^T$, we obtain
\begin{align*}
	& G(j\omega) + G^T(- j \omega)\\
= & \frac{ 4 \left( ( A_{l 1} + \text{Re}[R_{l1}])\text{Re}[d_l] - \text{Im}[d_l] \text{Im}[R_{l1}] \right) \omega^2 }{(|d_l|^2 - \omega^2)^2 + 4 \text{Re}[d_l]^2 \omega^2}\\
& + \frac{ 4 |d_l|^2 \left( ( A_{l 1} + \text{Re}[R_{l1}] ) \text{Re}[d_l] + \text{Im}[d_l] \text{Im}[R_{l1}] \right) }{(|d_l|^2 - \omega^2)^2 + 4 \text{Re}[d_l]^2 \omega^2}
\end{align*}
where the denominator is positive. Then, $G(j\omega) + G^T(- j \omega) \geq 0$ if the two numerators are positive semidefinite,  i.e.,
\begin{align}\label{eq: Al 1}
\begin{array}{rl}
	A_{l 1} + \text{Re}[R_{l1}] \pm \frac{\text{Im}[d_l] \text{Im}[R_{l1}]}{\text{Re}[d_l]} \geq 0.
\end{array}
\end{align}
When $k_l \geq 2$, we let $H(s) = \frac{R_{l k}}{(s + d_l)^{k}} + \frac{R_{l k}^*}{\left(s + d_l^*\right)^{k}} + \frac{A_{lk}}{s + d_l} + \frac{A_{lk}}{s + d_l^*}$. Note that $d_l = d_l^*$ and $R_{l k} = R_{l k}^*$ if $d_l \in \mathbb{R}$. Then,
\begin{align*}
& H(j \omega) + H^T( - j \omega)\\
=&  \frac{ R_{lk} + A_{lk} (j \omega + d_l )^{k - 1 }}{ (j \omega + d_l )^{k} } +  \frac{ R_{lk}^{H} + A_{lk} ( -j \omega + d_l^* )^{k - 1 }}{ ( - j \omega + d_l^* )^{k} } \\
& +  \frac{ R_{lk}^* + A_{lk} (j \omega + d_l^* )^{k - 1 }}{ (j \omega + d_l^* )^{k} }  +  \frac{ R_{lk}^T + A_{lk} ( -j \omega + d_l )^{k - 1 }}{ ( - j \omega + d_l )^{k} } \\
=& \frac{ R_{lk} (- j \omega + d_l^* )^{k} + R_{lk}^H ( j \omega + d_l )^{k}}{ ( j \omega + d_l)^{k} ( - j \omega + d_l^* )^{k} }\\
& + \hspace{-1mm}  \frac{ A_{lk} \hspace{-0.5mm}  \left[ (j \omega \hspace{-0.5mm} + \hspace{-0.5mm} d_l)^{k-1} (- \hspace{-0.5mm}  j \omega + d_l^*)^{k}  \hspace{-0.5mm}  + \hspace{-0.5mm}   ( - \hspace{-0.5mm}  j \omega \hspace{-0.5mm}  + \hspace{-0.5mm}  d_l^* )^{k-1} ( j \omega \hspace{-0.5mm} + \hspace{-0.5mm}  d_l)^{k} \right] }{{ ( j \omega \hspace{-0.5mm}  + \hspace{-0.5mm}  d_l)^{k} ( - j \omega \hspace{-0.5mm}  + \hspace{-0.5mm}  d_l^* )^{k} }} \\
& + \hspace{-1mm}  \frac{ R_{lk}^* (- j \omega + d_l )^{k} + R_{lk}^T ( j \omega + d_l^* )^{k} } { ( j \omega + d_l^*)^{k} ( - j \omega + d_l )^{k} }\\
& + \hspace{-1mm}  \frac{A_{lk} \hspace{-0.5mm}  \left[ ( j \omega \hspace{-0.5mm}  + \hspace{-0.5mm}  d_l^* )^{k-1} (- \hspace{-0.5mm}  j \omega + d_l)^{k}  \hspace{-0.5mm}  + \hspace{-0.5mm}  ( - \hspace{-0.5mm}  j \omega \hspace{-0.5mm}  + \hspace{-0.5mm}  d_l )^{k-1} ( j \omega \hspace{-0.5mm}  + \hspace{-0.5mm}  d_l^* )^{k} \right]} { ( j \omega + d_l^*)^{k} ( - \hspace{-0.5mm}  j \omega \hspace{-0.5mm} + \hspace{-0.5mm}  d_l )^{k} } \\
=&  \frac{ \cos k \bar{\theta} (R_{l k} + R_{l k}^{H}) \hspace{-0.5mm}  + \hspace{-0.5mm}  j \sin k \bar{\theta} (R_{l k}^{H} \hspace{-0.5mm}  - \hspace{-0.5mm}  R_{l k} ) \hspace{-0.5mm} + \hspace{-0.5mm}  2 A_{lk} \bar{r}^{k - 2} \text{Re}[d_l]}{\bar{r}^{k}}\\
& + \hspace{-1mm}  \frac{ \cos k \tilde{\theta} (R_{l k}^* \hspace{-0.5mm} + \hspace{-0.5mm}  R_{l k}^{T}) \hspace{-0.5mm} + \hspace{-0.5mm}  j \sin k \tilde{\theta} (R_{l k}^{T} \hspace{-0.5mm} - \hspace{-0.5mm}  R_{l k}^* ) \hspace{-0.5mm}  + \hspace{-0.5mm}  2 A_{lk} \tilde{r}^{k - 2} \text{Re}[d_l]}{\tilde{r}^{k}}
\end{align*}
where $\bar{r}$, $\bar{\theta}$ and $\tilde{r}$, $\tilde{\theta}$ are polar coordinates of $j \omega + d_l$ and $j\omega + d_l^*$, respectively. $H(j \omega) + H^T(-j \omega) \geq 0$ if the two numerators are positive semidefinite, i.e.,
\begin{equation}\label{eq:Al 2}
\begin{array}{rl}
	A_{lk} \geq & \left( \left\| {R_{lk} + R_{lk}^{H}} \right\| + \left\| {R_{lk}^{H} - R_{lk}} \right\|  \right) \frac{I}{ 2 \text{Re}[d_l]^{k-1}} \\
	A_{lk} \geq & \left( \left\| {R_{lk}^* + R_{lk}^{T}} \right\| + \left\| {R_{lk}^{T} - R_{lk}^*} \right\|  \right) \frac{I}{ 2 \text{Re}[d_l]^{k-1}}.
\end{array}
\end{equation}
Let $A_l = \sum_{k = 1}^{k_l} A_{l k}$, then summing up \eqref{eq: Al 1}, \eqref{eq:Al 2} for $k = 1,\ldots, k_l$ leads to \eqref{eq: PFC on MIMO}.
\end{pf}
We can observe that \eqref{eq: PFC on MIMO} is satisfied with some positive definite real matrix $A_l$.

\begin{rem}
If $R_{l1} \neq R_{l1}^T$ for some $l$ in \Cref{theorem: mimo passivation}, we can add a stable PFC $\frac{R_{l1}^T}{s + d_l}$ in advance to guarantee this precondition.
\end{rem}

\begin{rem}
We do not passivate terms with zero or pure imaginary poles, i.e., $A_l = 0$ if $\text{Re}[d_l] = 0$ because for IFP systems, $R_{l} \geq 0$ if $\text{Re}[d_l] = 0$.
\end{rem}

Similar results can be obtained for SISO linear system as a special case. Let $P_i$ be a SISO linear IFP system with transfer function $P_i(s)$. Then it can be decomposed into partial fractions as
\begin{align}\label{eq: transfer function siso}
P_i(s)  \hspace{-1mm} = \hspace{-1mm} \sum_{l=1}^{n} \left(\frac{c_{l1}}{s + d_l} + \frac{c_{l2}}{(s + d_l)^{2}} + \cdots + \frac{c_{lk_{l}}}{(s + d_l)^{k_l}} \right)
\end{align}
where $c_{lk} \in \mathbb{C}$ if $d_l \in \mathbb{C}$.
\begin{cor}\label{theorem: siso passivation}
The \textsc{SISO} \textsc{IFP} system given by
$P_i(s) = \sum_{l=1}^{n} \left(\frac{c_{l1}}{s + d_l} + \frac{c_{l2}}{(s + d_l)^{2}} + \cdots + \frac{c_{l k_{l}}}{(s + d_l)^{k_l}} \right)$ is input feedforward passivated by $C_i(s) = \sum_{l = 1}^{n} \frac{a_l}{s + d_l}$, where
\begin{align}
a_l \geq \frac{\left| \text{Im}[c_{l1}] \text{Im}[d_l] \right|}{\text{Re}[d_l]} + \sum_{k = 1}^{k_l} \frac{|c_{lk}|}{ \left( \text{Re} [d_l] \right)^{k-1}}
\end{align}
if $\text{Re}[d_l] > 0$ and $a_l = 0$, if $\text{Re}[d_l] = 0$.
\end{cor}

The compensator proposed in this work differs from the phase-lead compensators in \cite{yamashita2020passivity,li2020smooth} in that $d_l$ can be complex here.
It is noteworthy that a PFC is designed in \cite{kim2011passification,kim2016design} to render a SISO LTI plant minimum phase with relative degree one, which can be understood as a process of passivation when the plant is stable. Compared with \cite{kim2011passification,kim2016design}, our design method is more straightforward and can be carried out on MIMO linear systems.

%
\subsection*{Example 2}
\begin{figure*}[htbp]
\centering
\subfigure[ $C(a,s) = \frac{a}{(s + 0.5)}$,  $a =~1, ~2$, and $P(s) = \frac{1}{(s + 0.5)^2}$.]{\includegraphics[width = 0.40\linewidth]{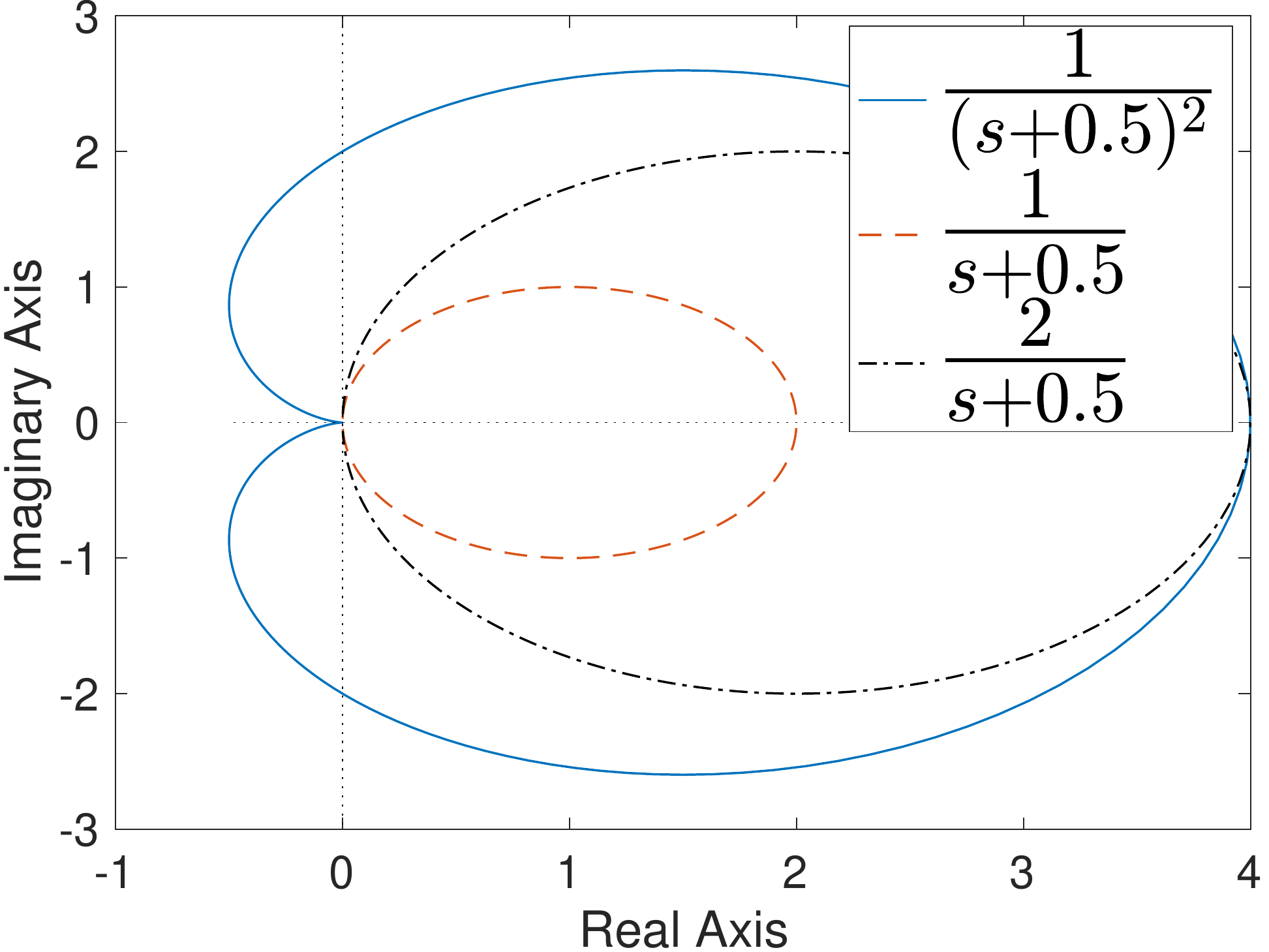} \label{fig: nyquist_H}}
\subfigure[$G(a,s) = \frac{1}{(s + 0.5)^2} + \frac{a}{s+ 0.5}$, $a = 1,~2,~3$ respectively.]{\includegraphics[width = 0.40\linewidth]{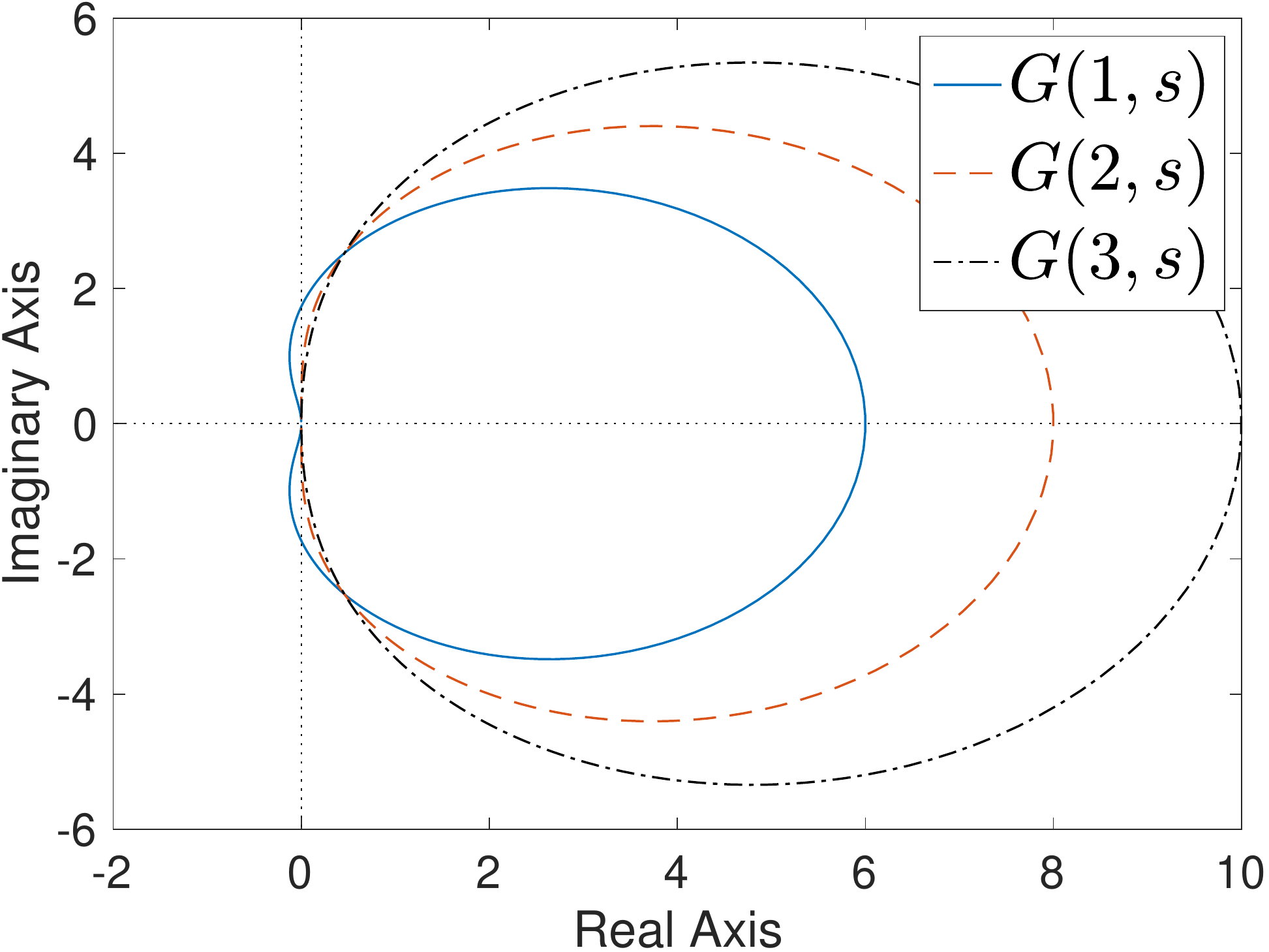} \label{fig: nyquist_P+C}}
\caption{Nyquist plots of $P(s)$, $C(a,s)$ and $G(s)$.}
\end{figure*}

Consider a SISO linear system with transfer function $P(s) = \frac{1}{(s + 0.5)^2}$, and a compensator $C(a,s) = \frac{a}{s + 0.5}$. The Nyquist plot of $P(s)$, $C(1,s)$, and $C(2,s)$ are shown in \Cref{fig: nyquist_H}. $P(s)$ is non-passive since it is not in the closed right-half plane while $C(a, s)$ is passive for any $a > 0$.
The Nyquist plot of $G(a,s) = P(s) + C(s)$ with $a = 1,~2, ~3$, respectively, are shown in \Cref{fig: nyquist_P+C}. It can be observed that the parallel interconnected system $G(a, s)$ is passive if and only if $a \geq 2$.


%



\subsection{Passivation of nonlinear input feedforward passive systems}
It is in general difficult to carry out frequency-wise analysis when system $P$ is nonlinear. In this case, it would be convenient to measure the passivity of a nonlinear system using passivity indices. The IFP index characterizes the shifting from a passive system in the frequency domain.
\begin{cor}
	A nonlinear \textsc{IFP($-\nu$)} system $P_i$ is input feedforward passivated by a compensator $C_i$ that is \textsc{IFP($\nu$)}, where $\nu > 0$.
\end{cor}
\begin{pf}
The sum of the storage functions of $P_i$ and $C_i$ is denoted by $V \geq 0$ and satisfies that $\dot{V} \leq y_1^T u_1 + \nu u_1^T u_1 + y_c^T u_c - \nu u_c^T u_c = \tilde{y}_1^T u_1$ since $u_1 = u_c$.
\end{pf}
To passivate an IFP($-\nu$) system with $\nu > 0$, compensator $C$ is required to be IFP($\nu$). Then the output $y_c$ of $C$ contains a direct feedthrough term of input, i.e., $D_c \neq 0$ \cite[Corollary 2.40]{sepulchre2012constructive}.
The passivation of IFP systems using PFC leads to fully distributed control for output synchronization, since the compensator $C_i$ for passivation is added locally to each system $P_i$, and does not require any graph information to design global parameters.


\subsection*{Example 3}\label{example 3}
We give an implicit application to show \icl{that synchronization is achieved via} fully distributed control and reveal \icl{how} some works are unified via the concept of PFC.
Consider the modified PI algorithm \cite{kia2015distributed,li2020input},
\begin{equation}\label{eq: ifp algorithm for quadratic programming}
\begin{array}{rl}
	\dot{x}_i = -\nabla f_i(x_i) - \lambda_i + u_i,
	\quad
	\dot{\lambda}_i = -u_i,
	\quad
	y_i = x_i
	\end{array}
\end{equation}
which is used to solve the distributed optimization problem $\min_{x \in \mathbb{R}} \sum_{i = 1}^{N} f_i(x)$. The input can take the diffusive couplings $u_i = \sigma \sum_{k =1}^{N} a_{ik} (y_k - y_i)$, with $\sigma > 0$ to be decided.
Let us consider quadratic convex functions $f_i(x_i)$ here. Then the transfer function for linear system \eqref{eq: ifp algorithm for quadratic programming} is
\[
\begin{array}{ll}
	G(s) = \frac{s + 1}{s\left( s + \nabla^2 f_i(x_i) \right)}
	 = \frac{ \left(\nabla^2 f_i(x_i) \right)^{-1} }{s} + \frac{1 - \left(\nabla^2 f_i(x_i) \right)^{-1} }{s + \nabla^2 f_i(x_i)}.
\end{array}
\]
When $\nabla^2 f_i (x_i) < 1$,  the system is passivity-short. \icl{Synchronization of the $y_i$ can be achieved by appropriately choosing the coupling gain $\sigma$ for input $u_i$, as proposed in \cite{qu2014modularized}; this, however, requires global graph information.
Following the analysis described in the previous sections, we} add the term $\frac{\left(\nabla^2 f_i(x_i) \right)^{-1} - 1 }{s + \nabla^2 f_i(x_i)}$ to $G(s)$, \icl{\il{which} is based on only the local agent dynamics}. The resulting system is passive, and becomes
\begin{equation}
\begin{array}{rl}
	\dot{x}_i =  \left(\nabla^2 f_i(x_i) \right)^{-1} u_i, \quad
	y_i =  u_i
\end{array}
\end{equation}
in the time domain, which is exactly the Zero-Gradient-Sum (ZGS) algorithm \cite{lu2012zero}, if $u_i$ takes the diffusive couplings. The closed-loop system is stable for any $\sigma > 0$.
Note that $\sigma$ does not depend on any graph information and the controller is thus fully distributed.

\section{Feedback Effects of PFC on Communication Graphs}\label{section: pfc on OS graphs}
\begin{figure}
	\centering
	\includegraphics[width = .6\linewidth,keepaspectratio]{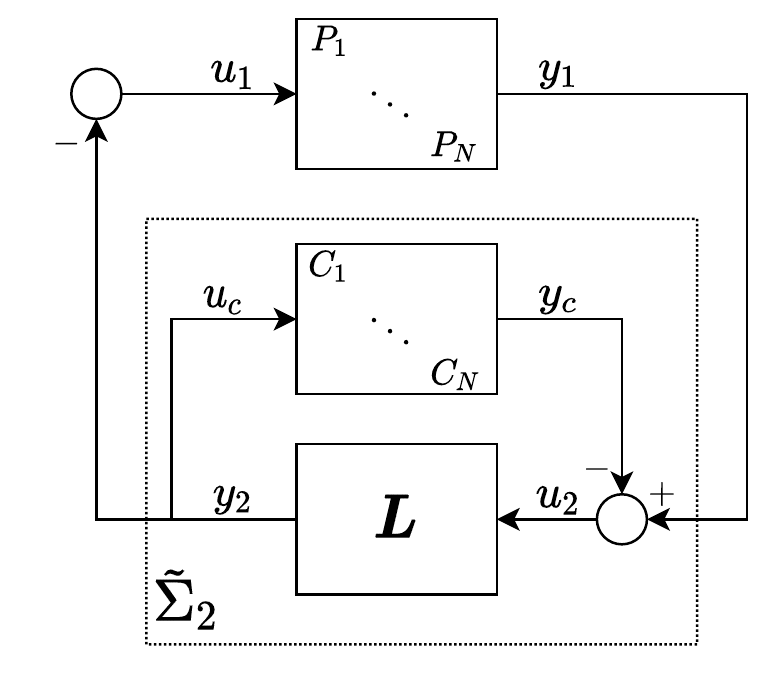}
	\caption{An equivalent block diagram representation of the addition of PFC to dynamics achieving output synchronization,  as in \Cref{fig: PFCtoOS}. The subsystem $\tilde{\Sigma}_2$ represents the feedback interconnection of $C$ and $\boldsymbol{L}$.}
	\label{fig: FPC to graph}
\end{figure}

The control diagram in \Cref{fig: PFCtoOS} is equivalent to \Cref{fig: FPC to graph}, where $C$ becomes a feedback compensator to $\boldsymbol{L}$.
In this section, we focus on the robustness effects of the PFC on communication graphs. The graph we consider in this section contains antagonistic interactions, represented by negative weight edges. We only assume that zero is a simple eigenvalue to $L$. In other words, $L$ may be indefinite.

\subsection{Convergence over signed weighted graphs}
We show that the use of PFC in output synchronization guarantees convergence of the closed-loop system when the graph is subject to negative weight edges.
First, let us characterize the passivity of the diffusive couplings under signed weighted graphs.
\begin{lem}\label{lemma: OFP of graph}
Suppose the graph is weight-balanced. The memoryless function $y = \boldsymbol{L} u$ is \textup{OFP($ - r$)} from $u$ to $y$, where $r$ is a constant satisfying
\begin{equation}\label{eq: OFP of graph}
	r L^T L + \frac{L+ L^T}{2} \geq 0.
\end{equation}
\end{lem}
\begin{pf}
A memoryless function $y = h(u)$ is output feedback passive if $y^T u + r y^T y \geq 0$, for some $r \in \mathbb{R}$. Substituting $y = \mathbf{L} u$ into the inequality, we obtain \eqref{eq: OFP of graph}.
Since $L^T L$ is positive semidefinite and has the same left and right null space as $L+L^T$ due to the graph being balanced, \eqref{eq: OFP of graph} has a feasible solution for some $r \in \mathbb{R}$.
\end{pf}
\begin{rem}
	Condition \eqref{eq: OFP of graph} is satisfied with some $r \leq 0$ when the graph does not have negative weight edges. In this case, $y = \mathbf{L} u$ is output strictly passive.
	However, when the graph contains negative weight edges, $L + L^T$ may not be positive semidefinite \cite{zelazo2014definiteness}. In this case, it is required that $r > 0$, and the mapping $y = \mathbf{L} u$ is OFP with a shortage of passivity.
	For unbalanced digraphs, a condition similar to \eqref{eq: OFP of graph} can be derived to show that $y = \boldsymbol{L} u$ is dissipative, which is not discussed in this work for simplicity.
\end{rem}

Consider the output synchronization of $N$ agents described by \eqref{eq: affine nonlinear subsystem} over a signed weighted digraph, as shown in \Cref{fig: PFCtoOS}.
Suppose that $C$ is passive with OFP index $\rho \geq 0$ and IFP index $\nu >0$. Then, $D_c$ satisfies $\frac{D_c^T + D_c}{2} \geq \nu I$ \cite{kottenstette2014relationships}.
Next, we show that the closed-loop system maintains passivity and reaches output synchronization.

\begin{thm}\label{theorem: convergence under antagonistic interaction}
Suppose each agent $P_i$ is passive, the signed weighted graph $\mathcal{G}$ is weight-balanced, strongly connected, and $0$ is a simple eigenvalue of $L$. Then, the closed-loop system in \Cref{fig: PFCtoOS} represented by \eqref{eq: affine nonlinear subsystem}, \eqref{eq: sigma_2 takes laplacian}, \icl{\eqref{eq:PFC_stable},} \eqref{eq: pfc state model}, \eqref{eq: pfc interconnection rule} is passive if $C$ is \textup{IFP}($\nu$), where $ \nu > r$, and $r$ is defined in \eqref{eq: OFP of graph}. The agents will also asymptotically achieve output synchronization.
\end{thm}
\begin{pf}
By \Cref{lemma: OFP of graph}, the mapping $y_2 = \boldsymbol{L} u_2$ is OFP($-r$).
Note that \Cref{fig: FPC to graph} is equivalent to \Cref{fig: PFCtoOS}.
System $\tilde{\Sigma}_2$ in \Cref{fig: FPC to graph} is output strictly passive from $u_2$ to $y_2$, by \Cref{lem:OFP}.
Then, the feedback interconnection of \il{the} two passive components $P$ and $\tilde{\Sigma}_2$ remains passive.
Let the Lyapunov function $V$ be the sum of \il{the} storage functions of $\Sigma_1$ and $\tilde{\Sigma}_2$. Since there is \il{no} external input, we have $\dot{V} \leq - \epsilon y_2^T y_2$, where $\epsilon = \nu - r > 0$. Since $V$ is positive definite and radially unbounded, the level set $\Omega=\{x:V(x)\leq \varepsilon \}$ for some $\varepsilon >0$ is a compact positively invariant set that can include any initial condition \icl{$x_0$} by \icl{choosing $\varepsilon = V(x_0)$}. It then follows from \il{LaSalle's invariance} principle that for any initial condition in $\Omega$ the trajectories converge to the largest invariant set $F\subset\Omega$ where $\dot{V} = 0$, i.e. $y_2 = \mathbf{0}$ in this set.
\icl{This therefore implies that for any initial condition we have $y_2 \to \mathbf{0}$ as $t\to\infty$.  Since $u_c=y_2$ we also have $u_c\to \mathbf{0}$, and the converging-input converging-output property of the PFC $C$ (property \eqref{eq:PFC_stable}) implies that $y_c \to \mathbf{0}$.
We now use the fact that $y_2 = \mathbf{L} (y_1 - y_c )$. Since $y_c \to \mathbf{0}$ and $y_2 \to \mathbf{0}$ we deduce that $\mathbf{L} y_1\to \mathbf{0}$, which implies $|y_{1i} - y_{1k}| \to 0$, $\forall{i,k}$, from the strong connectivity of the graph.}
\end{pf}

\begin{rem}
The graph Laplacian in this work follows from the definition $L := \text{diag}\{d_{in}^{1},\ldots, d_{in}^{N} \} - \mathcal{A}$, without taking absolute values in the diagonal terms, which avoids the bipartite consensus behavior as in \cite{altafini2012consensus}. In this case, some diagonal elements may even be negative, rendering negative eigenvalues, which is a case where convergence has not been shown in the literature \cite{altafini2012consensus,zelazo2014definiteness}. By utilizing the PFC technique, the positive semidefinite requirement is relaxed.
However, the PFC only helps improve convergence performance but does not affect the synchronization results. If zero is not a simple eigenvalue of $L$, clustering consensus may appear \cite{zelazo2014definiteness}.
\end{rem}

\subsection{PD control and practical derivative control}
Let the compensator be $C(s) = d_c \cdot I_{mN}$, which is a positive definite matrix that is IFP($- d_c$).
Then, the PFC simply \il{becomes derivative} feedback control. We let each agent $P_i$ be a simple integrator, i.e., $P_i = \frac{1}{s} I_{m}$. Then, the overall closed-loop system reduces to the proportional-derivative (PD) control for first-order consensus
\begin{equation}\label{eq: pd for consensus}
	\dot{x} = - \boldsymbol{L} x - d_c \boldsymbol{L} \dot{x}
\end{equation}
where $x = (x_1^T, \ldots, x_N^T)$ and $x_i$ is the local state of agent $i$. \Cref{theorem: convergence under antagonistic interaction} holds for sufficiently large $d_c$.
Therefore, a properly designed PD controller solves the synchronization problem under signed weighted graphs, if zero is a simple eigenvalue for $L$.

The derivative information is a prediction of future values. Hence, the real-time derivative cannot be obtained in practice. Nevertheless, we can realize the derivative control by taking the limit of a first-order low pass filter. For example, the derivative control in \eqref{eq: pd for consensus} can be realized by the compensator $C(s) = \underset{{\tau \rightarrow 0^{+}}}{\lim} \frac{d_c I}{\tau s + 1}$. In practice, each local compensator may be designed locally by selecting $\nu_i \geq \nu$, $\forall i \in \mathcal{N}$, where $\nu$ is the defined in \Cref{theorem: convergence under antagonistic interaction}.
Moreover, by adopting compensators with different internal dynamics, numerous new protocols can be designed. There are more benefits of adding this extra compensator. For example, it is found in \cite{ma2019exact} that the delay margin for consensus has been enlarged under the PD controller.

\subsection*{Example 4}
\begin{figure}[htbp]
	\centering
	\includegraphics[width = 0.4\linewidth]{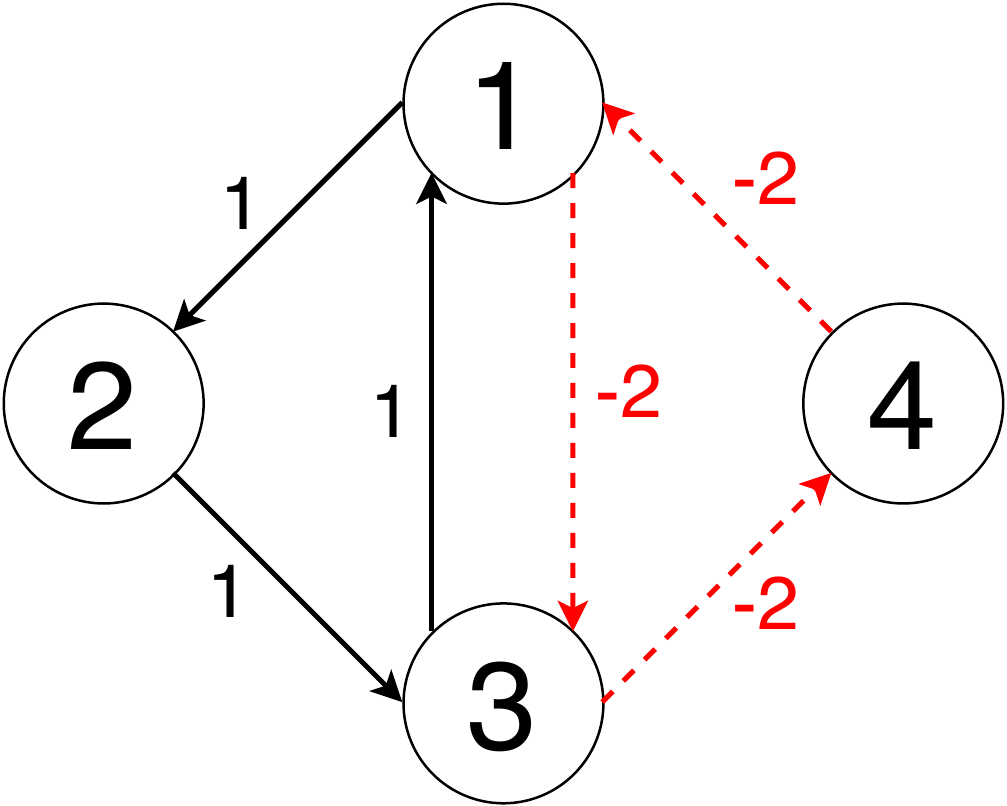}
	\caption{A balanced signed weighted communication digraph.}
	\label{fig: signed_weighted_graph}
\end{figure}

\begin{figure}[htbp]
\centering
\subfigure[Trajectories of $y_i$ without PFC.]{
	\includegraphics[width = 0.46\linewidth]{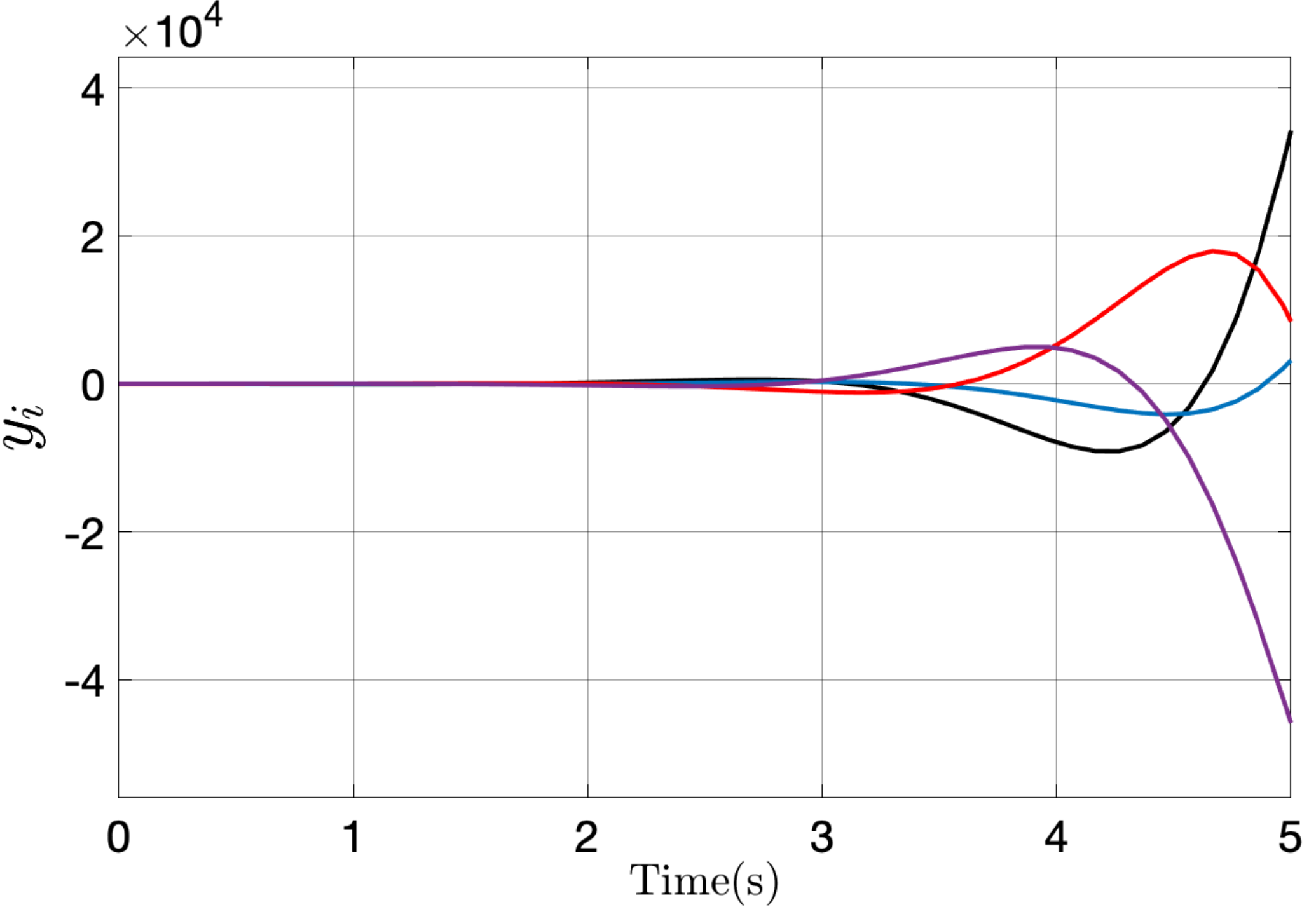}
	\label{fig: synchronization without PFC}}
\subfigure[Trajectories of $y_i$ with PFC.]{
	\includegraphics[width = 0.46\linewidth]{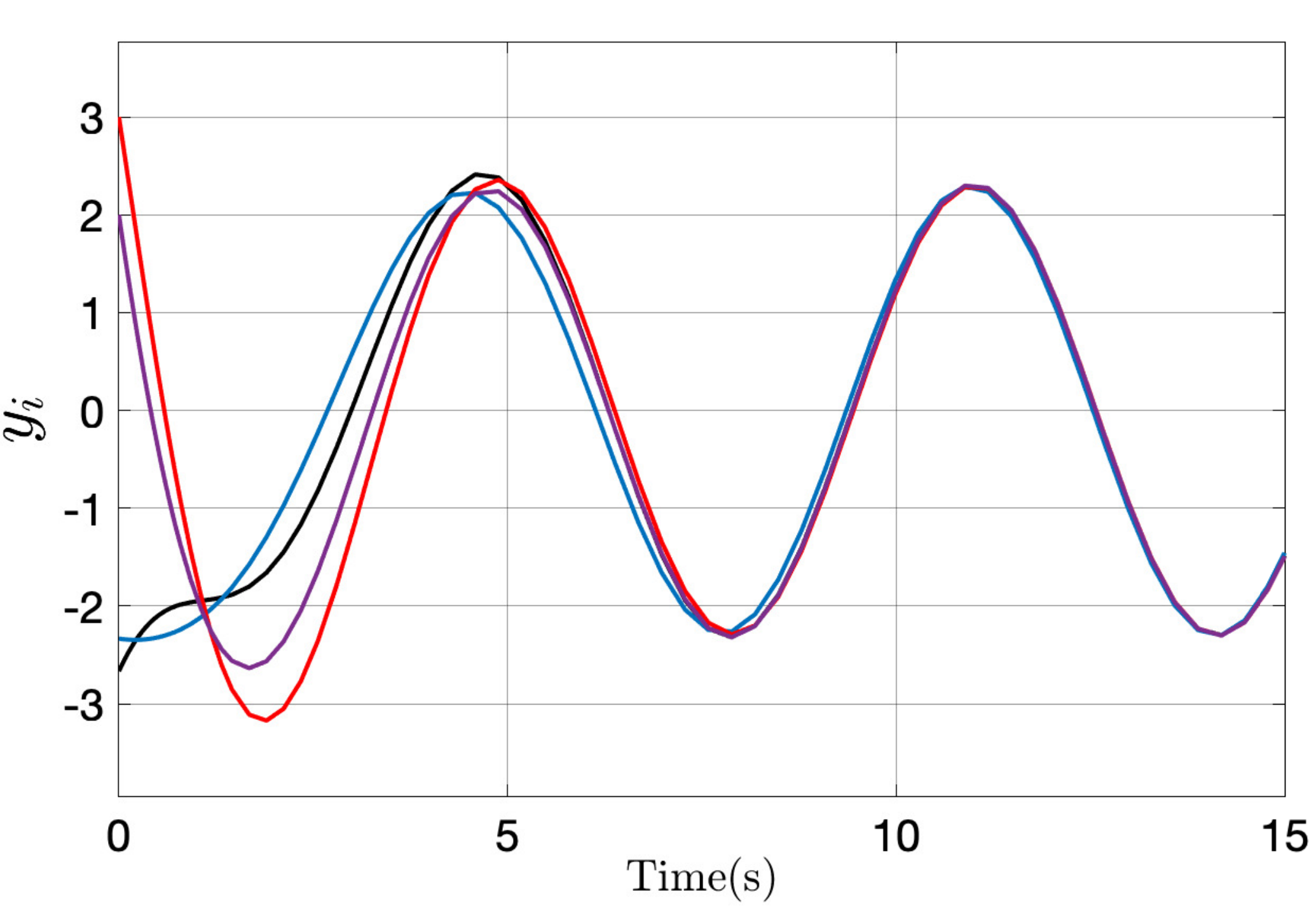}
	\label{fig: synchronization with PFC}
	}
	\caption{Performance of output synchronization represented by the closed-loop system \eqref{eq:example4} with and without PFC.}
	\label{fig: synchronization with and without PFC}
\end{figure}

Consider output synchronization of four harmonic oscillators represented by $P_i(s) = \frac{s}{s^2 + 1}$. Each oscillator is passive, then $P = \bigoplus_{i= 1}^{4} P_i$ is passive.
The communication graph is weight-balanced with negative weight edges, as shown in \Cref{fig: signed_weighted_graph}. The Laplacian is then given by $L = \begin{bmatrix} -1 & 0 & -1 & 2 \\ -1  & 1 & 0 & 0 \\ 2 & -1 & -1 & 0\\ 0 & 0 & 2 & -2\end{bmatrix}$, which is indefinite. The mapping $y = L u$ is OFP($-r$), where $r = 0.5$ by \Cref{lemma: OFP of graph}.
Let $C(s) = I$. The matrix $I + LC$ is nonsingular, so the overall system represented by \Cref{fig: FPC to graph} is well-posed.
The overall closed-loop system can be written as
\begin{subequations}\label{eq:example4}
\begin{align}
\dot{x}_i = &
\begin{bmatrix}
0 & -1\\ 1 & 0
\end{bmatrix}
x_i +
\begin{bmatrix}
1 \\ 0
\end{bmatrix}
u_i,
~
y_{i} =
\begin{bmatrix}
1 & 0
\end{bmatrix}
x_i,  ~i = 1, \ldots, 4,\\
u =&  - \left( I + L C \right)^{-1} L y.
\end{align}
\end{subequations}
We run the example with $C = \mathbf{0}$ (no addition of PFC), and $C = I$ on Simulink.
The trajectories of $y_i(t)$ for agents are shown in \Cref{fig: synchronization with and without PFC}. It can be observed that diffusive couplings cannot synchronize the outputs. The outputs blow to infinity due to antagonistic interactions.
On the other hand, the PFC provides extra derivative information that helps synchronization the outputs.

%

\section{Conclusion}\label{section: conclusion}
This work has investigated the effects of PFC on output synchronization from the perspective of passivity. A systematic method has been designed to passivate input feedforward passive systems. Moreover, it has been shown that the addition of PFC can solve output synchronization over signed weighted graphs with indefinite Laplacian.
There are also several aspects in this work that deserve further investigation in the future.
For instance, the communication graph can be extended to unbalanced cases; it is of interest to investigate the possibility of passivating nonlinear dynamics and indefinite Laplacians using a PFC without direct feedthrough terms.

\section*{CRediT authorship contribution statement}
\textbf{Mengmou~Li:} Conceptualization, Formal analysis, Methodology, Investigation, Software, Writing - review \& editing.
\textbf{Ioannis~Lestas:} Funding acquisition, Supervision, Validation, Writing - review \& editing.
\textbf{Li~Qiu:} Methodology, Funding acquisition, Supervision, Validation,Writing - review \& editing.

\section*{Declaration of competing interest}
The authors declare that they have no known competing financial interests or personal relationships that could have appeared to influence the work reported in this paper.

\section*{Acknowledgment}
This work was supported by the Hong Kong Research Grants Council under grant number 16201120, and by the European Research Council (ERC) under grant 679774.



  \bibliographystyle{elsarticle-num}
  \bibliography{References.bib}

\begin{thebibliography}{10}
\expandafter\ifx\csname url\endcsname\relax
  \def\url#1{\texttt{#1}}\fi
\expandafter\ifx\csname urlprefix\endcsname\relax\def\urlprefix{URL }\fi
\expandafter\ifx\csname href\endcsname\relax
  \def\href#1#2{#2} \def\path#1{#1}\fi

\bibitem{chopra2006passivity}
N.~Chopra, M.~W. Spong, Passivity-based control of multi-agent systems, in:
  Advances in robot control, Springer, 2006, pp. 107--134.

\bibitem{wieland2011internal}
P.~Wieland, R.~Sepulchre, F.~Allg{\"o}wer, An internal model principle is
  necessary and sufficient for linear output synchronization, Automatica 47~(5)
  (2011) 1068--1074.

\bibitem{wieland2013synchronous}
P.~Wieland, J.~Wu, F.~Allg{\"o}wer, On synchronous steady states and internal
  models of diffusively coupled systems, IEEE Transactions on Automatic Control
  58~(10) (2013) 2591--2602.

\bibitem{qu2014modularized}
Z.~Qu, M.~A. Simaan, Modularized design for cooperative control and
  plug-and-play operation of networked heterogeneous systems, Automatica 50~(9)
  (2014) 2405--2414.

\bibitem{isidori2014robust}
A.~Isidori, L.~Marconi, G.~Casadei, Robust output synchronization of a network
  of heterogeneous nonlinear agents via nonlinear regulation theory, IEEE
  Transactions on Automatic Control 59~(10) (2014) 2680--2691.

\bibitem{zhu2016general}
L.~Zhu, Z.~Chen, R.~H. Middleton, A general framework for robust output
  synchronization of heterogeneous nonlinear networked systems, IEEE
  Transactions on Automatic Control 61~(8) (2016) 2092--2107.

\bibitem{baldi2018output}
S.~Baldi, S.~Yuan, P.~Frasca, Output synchronization of unknown heterogeneous
  agents via distributed model reference adaptation, IEEE Transactions on
  Control of Network Systems 6~(2) (2018) 515--525.

\bibitem{bar1987parallel}
I.~Bar-Kana, On parallel feedforward and simplified adaptive control, IFAC
  Proceedings Volumes 20~(2) (1987) 99--104.

\bibitem{kim2016design}
H.~Kim, S.~Kim, J.~Back, H.~Shim, J.~H. Seo, Design of stable parallel
  feedforward compensator and its application to synchronization problem,
  Automatica 64 (2016) 208--216.

\bibitem{yamashita2020passivity}
S.~Yamashita, T.~Hatanaka, J.~Yamauchi, M.~Fujita, Passivity-based
  generalization of primal--dual dynamics for non-strictly convex cost
  functions, Automatica 112 (2020) 108712.

\bibitem{lombana2014distributed}
D.~A.~B. Lombana, M.~Di~Bernardo, Distributed {PID} control for consensus of
  homogeneous and heterogeneous networks, IEEE Transactions on Control of
  Network Systems 2~(2) (2014) 154--163.

\bibitem{ma2019exact}
D.~Ma, J.~Chen, R.~Lu, J.~Chen, Exact delay consensus margin of first-order
  agents under pid protocol, in: 2019 IEEE 58th Conference on Decision and
  Control (CDC), IEEE, 2019, pp. 54--59.

\bibitem{li2009consensus}
Z.~Li, Z.~Duan, G.~Chen, L.~Huang, Consensus of multiagent systems and
  synchronization of complex networks: A unified viewpoint, IEEE Transactions
  on Circuits and Systems I: Regular Papers 57~(1) (2009) 213--224.

\bibitem{kia2015distributed}
S.~S. Kia, J.~Cort{\'e}s, S.~Mart{\'\i}nez, Distributed convex optimization via
  continuous-time coordination algorithms with discrete-time communication,
  Automatica 55 (2015) 254--264.

\bibitem{li2020input}
M.~Li, G.~Chesi, Y.~Hong, Input-feedforward-passivity-based distributed
  optimization over jointly connected balanced digraphs, IEEE Transactions on
  Automatic Control 66~(9) (2021) 4117--4131.

\bibitem{lu2012zero}
J.~Lu, C.~Y. Tang, Zero-gradient-sum algorithms for distributed convex
  optimization: The continuous-time case, IEEE Transactions on Automatic
  Control 57~(9) (2012) 2348--2354.

\bibitem{khalil1996nonlinear}
H.~K. Khalil, Nonlinear systems, Prentice-Hall, New Jersey (2002).

\bibitem{sepulchre2012constructive}
R.~Sepulchre, M.~Jankovic, P.~V. Kokotovic, Constructive nonlinear control,
  Springer Science \& Business Media, 2012.

\bibitem{olfati2007consensus}
R.~Olfati-Saber, J.~A. Fax, R.~M. Murray, Consensus and cooperation in
  networked multi-agent systems, Proceedings of the IEEE 95~(1) (2007)
  215--233.

\bibitem{li2019consensus}
M.~Li, L.~Su, G.~Chesi, Consensus of heterogeneous multi-agent systems with
  diffusive couplings via passivity indices, IEEE Control Systems Letters 3~(2)
  (2019) 434--439.

\bibitem{altafini2012consensus}
C.~Altafini, Consensus problems on networks with antagonistic interactions,
  IEEE Transactions on Automatic Control 58~(4) (2012) 935--946.

\bibitem{zelazo2015robustness}
D.~Zelazo, M.~B{\"u}rger, On the robustness of uncertain consensus networks,
  IEEE Transactions on Control of Network Systems 4~(2) (2015) 170--178.

\bibitem{chen2016definiteness}
Y.~Chen, S.~Z. Khong, T.~T. Georgiou, On the definiteness of graph {L}aplacians
  with negative weights: Geometrical and passivity-based approaches, in: 2016
  American Control Conference (ACC), IEEE, 2016, pp. 2488--2493.

\bibitem{bronski2014spectral}
J.~C. Bronski, L.~DeVille, Spectral theory for dynamics on graphs containing
  attractive and repulsive interactions, SIAM Journal on Applied Mathematics
  74~(1) (2014) 83--105.

\bibitem{chen2020spectral}
W.~Chen, D.~Wang, J.~Liu, Y.~Chen, S.~Z. Khong, T.~Ba{\c{s}}ar, K.~H.
  Johansson, L.~Qiu, On spectral properties of signed {L}aplacians with
  connections to eventual positivity, IEEE Transactions on Automatic Control
  66~(5) (2021) 2177--2190.

\bibitem{zelazo2014definiteness}
D.~Zelazo, M.~B{\"u}rger, On the definiteness of the weighted {L}aplacian and
  its connection to effective resistance, in: 53rd IEEE Conference on Decision
  and Control, IEEE, 2014, pp. 2895--2900.

\bibitem{mukherjee2018robustness}
D.~Mukherjee, D.~Zelazo, Robustness of consensus over weighted digraphs, IEEE
  Transactions on Network Science and Engineering 6~(4) (2018) 657--670.

\bibitem{song2017network}
Y.~Song, D.~J. Hill, T.~Liu, Network-based analysis of small-disturbance angle
  stability of power systems, IEEE Transactions on Control of Network Systems
  5~(3) (2017) 901--912.

\bibitem{ding2017impact}
T.~Ding, R.~Bo, Y.~Yang, F.~Blaabjerg, Impact of negative reactance on
  definiteness of b-matrix and feasibility of dc power flow, IEEE Transactions
  on Smart Grid 10~(2) (2017) 1725--1734.

\bibitem{vidyasagar1981input}
M.~Vidyasagar, Input-output analysis of large-scale interconnected systems:
  decomposition, well-posedness and stability, Springer, 1981.

\bibitem{xia2018control}
M.~Xia, A.~Rahnama, S.~Wang, P.~J. Antsaklis, Control design using passivation
  for stability and performance, IEEE Transactions on Automatic control 63~(9)
  (2018) 2987--2993.

\bibitem{bao2007process}
J.~Bao, P.~L. Lee, Process control: the passive systems approach, Springer
  Science \& Business Media, 2007.

\bibitem{li2020smooth}
M.~Li, S.~Yamashita, T.~Hatanaka, G.~Chesi, Smooth dynamics for distributed
  constrained optimization with heterogeneous delays, IEEE Control Systems
  Letters 4~(3) (2020) 626--631.

\bibitem{kim2011passification}
S.~Kim, H.~Kim, J.~Back, H.~Shim, J.~H. Seo, Passification of {SISO} {LTI}
  systems through a stable feedforward compensator, in: 2011 11th International
  Conference on Control, Automation and Systems, IEEE, 2011, pp. 107--111.

\bibitem{kottenstette2014relationships}
N.~Kottenstette, M.~J. McCourt, M.~Xia, V.~Gupta, P.~J. Antsaklis, On
  relationships among passivity, positive realness, and dissipativity in linear
  systems, Automatica 50~(4) (2014) 1003--1016.

\end{thebibliography}


%
%
%
\end{document}